\documentclass[12pt]{iopart}
\newcommand{\lp}{\left(}
\newcommand{\rp}{\right)}
\newcommand{\lb}{\left[}
\newcommand{\rb}{\right]}
\newcommand{\lbar}{\left|}
\newcommand{\rbar}{\right|}
\newcommand{\non}{\nonumber}
\newcommand{\dopr}{\prime\prime}
\newcommand{\ndi}{\dotlessi}
\newcommand{\ndj}{\dotlessj}
\newcommand{\intd}{\int\displaylimits}	
\newcommand{\sumd}{\sum\displaylimits}
\newcommand{\rv}{{\bf r}}                                               
\newcommand{\rvp}{{\bf r}^{\prime}}                                               
\newcommand{\bfx}{{\bf x}}                                              
\newcommand{\bfxp}{{\bf x}^{\prime}}
\newcommand{\ov}{\overline{v}}                                          
\newcommand{\orr}{\overline{R}}                                         
\newcommand{\ootw}{\frac{1}{2}}                                         
\newcommand{\be}{\begin{equation}} 
\newcommand{\ee}{\end{equation}} 
\newcommand{\bea}{\begin{eqnarray}} 
\newcommand{\eea}{\end{eqnarray}} 
\newcommand{\bm}{\begin{mathletters}} 
\newcommand{\eml}{\end{mathletters}}

\newcommand{\of}{\frac{1}{4}}

\newcommand{\la}{\langle} 
\newcommand{\ra}{\rangle}

\usepackage{graphicx}							
\usepackage{tensor}							
\usepackage{dotlessi}                                                   
\usepackage[intlimits]{esint}                                           

\begin{document}

\title{Self--consistent description of emission processes in 
axially--symmetric nuclei}

\author{A Dumitrescu$^{1}$, D S Delion$^{1,2}$}

\address{$^{1}$ Horia Hulubei National Institute for R\&D in Physics and
         Nuclear Engineering, No. 30 Reactorului Street, Magurele, 
         077125, Ilfov, Romania}
\address{$^{2}$ Academy of Romanian Scientists, No. 3 Ilfov Street, 
         Bucharest, 050044, sector 5, Romania}
\ead{alexandru.dumitrescu@theory.nipne.ro}
\vspace{10pt}
\begin{indented}
\item[]July 2026
\end{indented}

\begin{abstract}
We present a theory of cluster emission processes in terms of 
proton ($\pi$) and neutron ($\nu$) single--particle (sp) degrees of
freedom within a self--consistent mean--field (SCMF) constructed from
a two--particle interaction having relative (rel) and center of mass
(com) terms centered on the nucler surface, the latter describing the 
interaction between the com of a pair of particles and the surface of an 
axially--symmetric nucleus. In this way the $\alpha$-clustering 
phenomenon becomes enhanced on the nuclear surface. We present 
applications for unstable nuclei that decay through the emission of 
$\alpha$--particles above $\tensor[^{100}]{\textrm{Sn}}{}$,
$\tensor[^{208}]{\textrm{Pb}}{}$ and within the actinidies series.
\end{abstract}

%
\vspace{2pc}
\newline
\newline
%
\noindent{\submitto{\JPG}}
%
%
%

\section{Introduction}
\label{introduction}

The theory of $\alpha$--decay was first established by Gamow 
\cite{Gam28} and independently by Gurney \& Condon \cite{Gur28}, 
representing the first theoretical framework for the probabilistic 
interpretation of quantum mechanics. Historical notes on radioactivity 
and a broad overview of the theoretical research and some experimental 
aspects investigated over the decades can be found in Refs. \cite{Lov98,
Del15,Del18,Dum22,Dum25}. Of particular interest for this work is the 
microscopic theory of $\alpha$--decay, particularly the clustering 
mechanism described in terms of antisymmetrized nucleonic degrees of 
freedom. Foundational work was carried out in the 1960s \cite{San62,
Man64,Pog69} based on earlier results of Lane \cite{Lan60}. The 
techniques for rigorous antisymmetrization were investigated by 
Fliessbach throughout the 1970s \cite{Fli75,Fli76,Fli77,Fli79}. In the 
1990s the problem of nucleonic clustering was studied extensively within
infinite nuclear matter \cite{Rop98}, while modern approaches to finite 
systems along with many interesting developments took place in the early 
2000s. We mention studies of the Hoyle state \cite{Toh01,Epe11}, 
approaches using energy density functionals \cite{Ebr12}, calculations 
for medium mass or heavy nuclei \cite{Das23,Bet12,Rop14} and novel 
methods as through nonlinear quantum dynamics \cite{Car21}. 

The present work continues investigations started at the spherical 
mean--field (MF) level by Delion and Liotta \cite{Del13} and developed
further for spherical systems in a self--consistent approach by 
Dumitrescu and Delion \cite{Dum23,Dum25}, the focus of the present
research being axially--symmetric nuclei. Our discussion is organized
as follows. Section \ref{theory} covers the formal developments of our
Cluster--Hartree--Fock--Bardeen--Cooper--Schrieffer (CHFBCS) theory for 
axially--symmetric clustered nucleonic systems, together with a 
calculation of the cluster formation amplitude using an
approximate antisymmetrization technique. Section \ref{numap} is
concerned with applications to open shell nuclei above 
$\tensor[^{100}]{\textrm{Sn}}{}$ and for decay chains starting in
$\tensor[^{208,210,212}]{\textrm{Pb}}{}$ and followed through into the
actinides region. Section \ref{conc} presents our conclusions.

\section{CHFBCS theory}
\label{theory}

In this section we generalize the approach of Ref. \cite{Dum25} to
axially--symmetric nuclei, presenting the corresponding CHFBCS
theory where the cluster is once more illustrated by an 
$\alpha$--particle. Our goal is to obtain simultaneously a good 
description of the ground--state (gs) properties of an unstable nucleus
together with a good estimation of its decay width in the 
$\alpha$--channel. Section \ref{sgi} details the properties of the 
Surface Gaussian Interaction (SGI) from which the SCMFs are constructed. 
Section \ref{hfeq} covers the formal derivation of the CHFBCS equations 
and section \ref{wid} details the calculation of the antisymmetrized
$\alpha$--particle formation amplitude in an axially--symmetric field.

Let us mentions in this context that the standard 
Hartree--Fock--Bogoliubov (HFB) approach describing deformed nuclei uses 
a cranking procedure on the quadrupole moment in order to induce 
deformation. In this case one uses the generalized quasiparticle 
representation involving all sp orbitals due to the non--diagonality of 
the quadrupole operator. We will induce deformation by a surface 
clustering term and therefore the quasiparticle representation needs 
only the Bardeen--Cooper--Schrieffer (BCS) treatment coupling sp states 
with opposite spin projections.

\subsection{Surface Gaussian Interaction}
\label{sgi}

The two--particle interaction is defined as a Wigner force 
\bea
\label{vsgi}
v\lp\rv_{1},\rv_{2}\rp=-\ov_{\tau}
e^{-\frac{\lbar\rv_{1}-\rv_{2}\rbar^{2}}{b^{2}}}\lb 1+x_{\tau}
e^{-\frac{\lp R-R_{0}\rp^{2}}{B^{2}}}\rb.
\eea

$\rv_{1,2}$ are the radial vectors of the interacting particles. 
$\ov_{\tau}$ is the interaction constant, depending on the isospin
$\tau=\pi,\nu$. $b$ and $B$ are interaction lengths for the rel and com
terms. $R$ is the com radius for the given pair
\bea
R=\ootw\sqrt{r_{1}^{2}+2r_{1}r_{2}\cos\theta_{12}+r_{2}^{2}}
\eea 
with $\theta_{12}$ the relative angle between the particles. $R_{0}$ is 
the clustering radius in the com direction
\bea
R_{0}=\orr_{0}\lb 1+\beta_{2} Y_{20}\lp\theta_{\textrm{com}}\rp\rb
\eea
in terms of a spherical radius $\orr_{0}$ and the quadrupole elongation
$\beta_{2}$. The direction of the com relative to the polar axis 
follows from
\bea
\cos\theta_{\textrm{com}}=\frac{r_{1}\gamma_{1}+r_{2}\gamma_{2}}{R}
\eea 
where in general $\gamma_{\ndi}=\cos\theta_{\ndi},~\ndi=1,2$, i.e. it is 
the cosine of the polar angle for particle $\ndi$. $x_{\tau}$ is a 
control parameter necessary in order for the interaction to reproduce 
the decay width. We will show later that the values of this parameter 
for different isospins are not independent and that physically they 
correspond to the amplitude of a surface pocket in the resulting SCMF. 
Because the second term of the interaction ultimately depends both on 
$\theta_{12}$ and $\theta_{\textrm{com}}$, one cannot use throughout the 
usual multipole expansion in terms of Legendre polynomials, but rather 
the more general version
\bea
v\lp\rv_{1},\rv_{2}\rp=\sumd_{LM}\frac{4\pi}{\hat{L}^{2}}
v_{L}\lp r_{1},r_{2}\rp Y_{LM}\lp\Omega_{1}\rp 
Y^{*}_{LM}\lp\Omega_{2}\rp
\eea
where $\Omega_{\ndi}$ is the solid angle of particle $\ndi$. The radial
form factors follow as
\bea
v_{L}\lp r_{1},r_{2}\rp=-\ov_{\tau}v_{L}^{\lp\textrm{rel}\rp}
\lp r_{1},r_{2}\rp \lb 1+x_{\tau}v_{L}^{\lp\textrm{com}\rp}
\lp r_{1},r_{2}\rp\rb.
\eea

As the rel term of the expansion depends only on $\theta_{12}$, its 
evaluation is entirely equivalent to the usual result given in terms of
Legendre polynomials
\bea
v_{L}^{\lp\textrm{rel}\rp}\lp r_{1},r_{2}\rp=\frac{\hat{L}^{2}}{2}
\intd_{-1}^{1}\mathrm{d}\gamma_{12}P_{L}\lp\gamma_{12}\rp
e^{-\frac{\lp r_{1}-r_{2}\rp^{2}-2r_{1}r_{2}\lp\gamma_{12}-1\rp}
{b^{2}}}.
\eea

The com term is more involved, integrating as
\bea
&~&v_{L}^{\lp\textrm{com}\rp}\lp r_{1},r_{2}\rp=\frac{\hat{L}^{4}}{4}
\iint_{-1}^{1}\mathrm{d}\gamma_{1}\mathrm{d}\gamma_{2}
P_{L}\lp\gamma_{1}\rp P_{L}\lp\gamma_{2}\rp\times\non\\
&~&e^{-\frac{r_{1}^{2}+r_{2}^{2}-2r_{1}r_{2}
\lb\sqrt{\lp 1-\gamma_{1}^{2}\rp\lp 1-\gamma_{2}^{2}\rp}
-\gamma_{1}\gamma_{2}\rb}{b^{2}}}\times\non\\
&~&\intd_{0}^{1}\mathrm{d}t e^{\frac{2r_{1}r_{2}
\sqrt{\lp 1-\gamma_{1}^{2}\rp\lp 1-\gamma_{2}^{2}\rp}}{b^{2}}
\lb\cos\lp 2\pi t\rp-1\rb}e^{-\frac{\lp R-R_{0}\rp^{2}}{B^{2}}}.
\eea

In this instance, we have scaled $\phi_{12}=2\pi t$, where $\phi_{12}$
is the phase difference between the particles in the equatorial plane. 
Thus, the angle $\theta_{12}$ in the expression of $R$ also evaluates
in terms of $\gamma_{1},\gamma_{2}$ and $\phi_{12}$. We note that in the 
instances where the com exponential can be neglected, the last integral 
in the com radial form factor is just the integral representation of a
modified Bessel function of order 0. 

\subsection{CHFBCS equations}
\label{hfeq}

The HF equations for an axially-symmetric nucleus, written within the
intrinsic frame of reference, are
\bea
-\frac{\hbar^{2}}{2m_{N}}\nabla^{2}\Phi_{m}\lp\bfx\rp+U\lp\bfx\rp
\Phi_{m}\lp\bfx\rp+W\lp\bfx,\bfxp\rp=\epsilon_{m}\Phi_{m}\lp\bfx\rp.
\eea

The Nilsson wave functions expand in partial Woods--Saxon (WS) spherical
waves
\bea
\Phi_{m}\lp\bfx\rp=\sumd_{\ell\ndj}\frac{g_{\ell\ndj m}\lp r\rp}{r}
\mathcal{Y}_{\ndj m}^{\lp\ell\ootw\rp}\lp\Omega,\chi\rp
\eea
having the total angular momentum $\ndj\ge m$, where $m$ is the 
intrisic projection of the Nilsson wave function. $\bfx$ represents
spatial and spin degrees of freedom while $\Omega$ is a solid angle
and $\chi$ a spinor of spin $\ootw$. $U\lp\bfx\rp$ is the direct
term while $W\lp\bfx,\bfxp\rp$ is the exchange term. Our previous
studies \cite{Dum23,Dum25} have shown that the exchange term of the HF
equations is negligible for the main purpose of our investigations, 
namely the calculation of a cluster formation amplitude, as its effects 
near the nuclear surface are minor. As such, we neglect it completely 
for the present study and focus only on the direct term.

When expanded in a spherical harmonic oscillator basis (ho), the radial
WS wave functions have the expressions
\bea
g_{\ell\ndj m}\lp r\rp=\sumd_{n}d_{\tau\epsilon\ell\ndj m}^{\lp n\rp}
\mathcal{R}_{n\ell}^{\lp\beta\rp}\lp r\rp
\eea
while the spin--orbit harmonics are given by
\bea
\mathcal{Y}_{\ndj m}^{\lp\ell\ootw\rp}\lp\Omega,\chi\rp=i^{\ell}
\lb Y_{\ell}\otimes\chi_{\ootw}\rb_{\ndj m}.
\eea
The additional quantum numbers involved are the sp energy $\epsilon$, 
the orbital angular momentum $\ell$ and the radial ho number $n$.
The wave functions depend upon the ho mass parameter $\beta=
\frac{m_{N}\omega}{\hbar}$, where $m_{N}$ is the average nucleon mass
and $\omega$ is the usual ho frequency
\bea
\hbar\omega=41\cdot A^{-\frac{1}{3}}~\textrm{MeV}
\eea
depending on the mass number $A$. The direct term is defined as
\bea
U\lp\bfx\rp=\intd_{\mathcal{D}^{\prime}}\textrm{d}^{3}\rvp 
v\lp\rv,\rvp\rp\rho\lp\bfxp\rp
\eea
in terms of the density
\bea
\rho\lp\bfxp\rp=\sumd_{m}v_{m}^{2}\Phi_{m}^{\dagger}\lp\bfxp\rp
\Phi_{m}\lp\bfxp\rp
\eea
where $v_{m}^{2}$ are the standard BCS occupation probabilities. At any
point of the integration domain, the density can be expanded in 
multipoles
\bea
\rho\lp\rv\rp=\sumd_{L}\frac{\sqrt{4\pi}}{\hat{L}}\rho_{L}\lp r\rp
Y_{L0}\lp\theta\rp
\eea
leading to the radial form factors
\bea
\rho_{L}\lp r\rp&=&\frac{\hat{L}^{2}}{4\pi}\sumd_{m}v_{m}^{2}
\sumd_{\ell\ndj\ell^{\prime}\ndj^{\prime}}\frac{1}{r^{2}}
g_{\ell\ndj m}\lp r\rp g_{\ell^{\prime}\ndj^{\prime} m}\lp r\rp
i^{\ell-\ell^{\prime}}\times\non\\
&~&\langle\ndj\ootw;L0|\ndj^{\prime}\ootw\rangle
\langle\ndj^{\prime}m;L0|\ndj m\rangle
\eea
involving the usual Clebsch--Gordan coefficients denoted by bra--ket
products. Similarly, the multipole expansion of the direct term is given 
by
\bea
U\lp\rv\rp=\sumd_{L}\frac{\sqrt{4\pi}}{\hat{L}}U_{L}\lp r\rp Y_{L0}
\lp\theta\rp
\eea 
with the radial form factors
\bea
\label{uuffs}
U_{L}\lp r\rp=\frac{4\pi}{\hat{L}^{2}}\intd_{0}^{\infty}\mathrm{d}
r^{\prime}r^{\prime 2}v_{L}\lp r,r^{\prime}\rp\rho_{L}\lp r^{\prime}\rp.
\eea

When the HF equations are integrated over $\intd_{\lp 4\pi\rp}
\mathrm{d}\Omega\lb\mathcal{Y}_{\ndj m}^{\lp\ell\ootw\rp}
\lp\Omega,\chi\rp\rb^{\dagger}$, the following set of coupled second 
order radial differential equations is obtained
\bea
-\frac{\hbar^{2}}{2m_{N}}\lb g_{\ell\ndj m}^{\dopr}\lp r\rp-
\frac{\ell\lp\ell+1\rp}{r^{2}}g_{\ell\ndj m}\lp r\rp\rb+\non\\
\sumd_{L}\frac{\sqrt{4\pi}}{\hat{L}\hat{\ndj}}U_{L}\lp r\rp
\sumd_{\ell^{\prime}\ndj^{\prime}}g_{\ell^{\prime}\ndj^{\prime} m}
\lp r\rp\langle\ndj\lbar\lbar Y_{L}\rbar\rbar\ndj^{\prime}\rangle
\langle\ndj^{\prime}m;L0|\ndj m\rangle=
\epsilon_{m}g_{\ell\ndj m}\lp r\rp
\eea 
where the coupling involves the reduced matrix element 
$\langle\ndj\lbar\lbar Y_{L}\rbar\rbar\ndj^{\prime}\rangle$ following
from the Wigner-Eckart theorem. Using the ho expansion of the WS wave 
functions, this system can be mapped onto the eigenvalues and 
eigenvectors problem \cite{Del10}
\bea
\sumd_{n^{\prime}\ell^{\prime}\ndj^{\prime}}
H^{\lp\beta\rp}_{n\ell\ndj,n^{\prime}\ell^{\prime}\ndj^{\prime}}
d^{\lp n^{\prime}\rp}_{\tau\epsilon\ell^{\prime}\ndj^{\prime}m}=
\epsilon_{m}d^{\lp n\rp}_{\tau\epsilon\ell\ndj m}
\eea
with the Hamiltonian matrix
\bea
H^{\lp\beta\rp}_{n\ell\ndj,n^{\prime}\ell^{\prime}\ndj^{\prime}}=
\bigg[\hbar\omega\lp 2n+\ell+\frac{3}{2}\rp\delta_{nn^{\prime}}+
\langle\beta n\ell\lbar U_{0}\lp r\rp\rbar
\beta n^{\prime}\ell\rangle-\non\\
\frac{\hbar\omega}{2}
\langle\beta n\ell\lbar\beta r^{2}\rbar\beta n^{\prime}\ell\rangle\bigg]
\delta_{\ell\ell^{\prime}}\delta_{\ndj\ndj^{\prime}}+
\langle\beta n\ell\lbar
V_{d}^{\lp\ell\ndj\ell^{\prime}\ndj^{\prime}\rp}\lp r\rp\rbar
\beta n^{\prime}\ell^{\prime}\rangle
\eea
given in terms of the monopole $U_{0}\lp r\rp$ and deformed potential
\bea
V_{d}^{\lp\ell\ndj\ell^{\prime}\ndj^{\prime}\rp}\lp r\rp=
\sumd_{L>0}\frac{\sqrt{4\pi}}{\hat{L}\hat{\ndj}}\langle
\ndj\lbar\lbar Y_{L}\rbar\rbar\ndj^{\prime}\rangle
\langle\ndj^{\prime}m;L0|\ndj m\rangle U_{L}\lp r\rp.
\eea

It is worth noting that such HF problems can also be treated in a 
Cartesian deformed ho basis, with many complex and interesting 
applications \cite{Dob21}.

\subsection{Decay width}
\label{wid}

The total $\alpha$-decay width can be derived from the continuity 
equation \cite{Del10}. The result is a sum of partial decay widths 
corresponding to different angular momenta $L_{\alpha}$ of the emitted 
cluster
\bea
\label{gam}
\Gamma=\sumd_{L_{\alpha}}\Gamma_{L_{\alpha}}=\sumd_{L_{\alpha}}\hbar 
v_{\alpha}\lbar N_{L_{\alpha}}\rbar^2
\eea
where $v_{\alpha}=\sqrt{\frac{2Q_{\alpha}}{\mu_{\alpha}}}$ is the 
asymptotic $\alpha$-particle velocity depending on the Q-value of the 
process $Q_{\alpha}$ and the reduced mass of the alpha-daughter system 
$\mu_{\alpha}$. The quantity $N_{L_{\alpha}}$ is called channel 
scattering amplitude and can be found by matching the internal and 
external components of the $\alpha$-particle wave function
\bea
\label{match}
f^{\lp\textrm{int}\rp}_{L_{\alpha}}\lp R_{\alpha}\rp=
f^{\lp\textrm{ext}\rp}_{L_{\alpha}}\lp R_{\alpha}\rp.
\eea
The external component is built as a superposition of fundamental 
solutions with the outgoing Coulomb--Hankel asymptotics
\bea
f^{\lp\textrm{ext}\rp}_{L_{\alpha}}\lp R_{\alpha}\rp=
\sumd_{L^{\prime}_{\alpha}}\mathcal{H}^{\lp+\rp}_{L_{\alpha}
L^{\prime}_{\alpha}}\lp R_{\alpha}\rp N_{L^{\prime}_{\alpha}}
\ \vtop{\halign{#\cr $\longrightarrow$\cr \hfil$\mskip -10mu 
\scriptstyle R_{\alpha}\rightarrow\infty$\hfil\cr}} \ 
H^{\lp+\rp}_{L_{\alpha}}\lp\chi,\kappa R_{\alpha}\rp
\eea
depending upon the Coulomb parameter $\chi=\frac{4Z_{D}}
{\hbar v_{\alpha}}$ and reduced radius $\kappa R_{\alpha}$
with $\kappa=\mu_{\alpha}v_{\alpha}$. The matrix of fundamental 
solutions $\mathcal{H}^{\lp+\rp}_{L_{\alpha}L^{\prime}_{\alpha}}$
can be found through backward integration, while the channel scattering 
amplitudes are given by inverting this matrix in Eq. (\ref{match})
\bea
N_{L_{\alpha}}=\sumd_{L^{\prime}_{\alpha}}\lb
\mathcal{H}^{\lp+\rp}_{L_{\alpha}L^{\prime}_{\alpha}}
\lp R_{\alpha}\rp\rb^{-1}
f^{\lp\textrm{int}\rp}_{L^{\prime}_{\alpha}}\lp R_{\alpha}\rp
\eea
Let us mention that a semiclassical approximation for the matrix of 
fundamental solutions in a pure Coulomb field is given by the Fr\"{o}man 
approach \cite{Fro57}. The internal function is called $\alpha$-particle 
formation amplitude and is given by the overlap between parent and 
antisymmetrized product of daughter and $\alpha$-particle wave functions
\bea
f^{\lp\textrm{int}\rp}\lp{\bf R}_{\alpha}\rp=\la\psi_{P}|
\mathcal{A}\lp\psi_{D}\psi_{\alpha}\rp\ra.
\eea
In several of our previous works we estimated this integral at distances 
beyond the geometrical contact radius, where the Pauli principle becomes 
less important and therefore we neglected the daughter-cluster 
antisymmetrization procedure. For a multipole component of the formation 
amplitude, the result becomes a superposition in terms of radial ho wave 
functions depending on radial and angular $\alpha$-particle quantum 
numbers
\bea
f^{\lp\textrm{int}\rp}_{L_{\alpha}}\lp R_{\alpha}\rp=\sumd_{N_{\alpha}}
W_{N_{\alpha}L_{\alpha}}
\mathcal{R}^{\lp 4\beta\rp}_{N_{\alpha}L_{\alpha}}\lp R_{\alpha}\rp.
\eea
The coefficients $W$ are given in Chapter 9.6.1 of Ref. \cite{Del10} in 
terms of the recoupling of Moshinsky symbols, the $uv$ products of BCS 
amplitudes and sp Nilsson expansion coefficients in the ho basis. In 
the present work it is necessary to compute the $\alpha$-particle 
formation amplitude at smaller distances, where the clustering component 
has its maximal value. We estimate the contribution of the 
antisymmetrization by considering the largest Slater determinant at the 
Fermi level in the BCS wave function. We consider the combinatorial 
daughter--$\alpha$--particle term used in Ref. \cite{Man64}
fully taking into account daughter-$\alpha$-particle antisymmetrization
inside the nucleus where $\rho\sim\rho_0$, and consider the value of 
unity in the limit of large distances, where the nuclear density 
vanishes. Thus
\bea
\label{asym}
\mathcal{F}_{L_{\alpha}}\lp R_{\alpha}\rp=
f^{\lp\textrm{int}\rp}_{L_{\alpha}}\lp R_{\alpha}\rp
\lb\lp C-1\rp\frac{\rho}{\rho_{0}}+1\rb^{\frac{1}{2}}
\eea
where
\bea
C=\lp\matrix{ Z_m \cr 2 }\rp
\lp\matrix{ N_m \cr 2 }\rp.
\eea
Here we considered in the antisymmetrization procedure the particles 
above the closest magic number
\bea
Z_{m}=Z-Z_{\textrm{magic}},~N_{m}=N-N_{\textrm{magic}}
\eea
and the standard density parametrisation
\bea
\frac{\rho}{\rho_0}=\frac{1}{1+\exp[(R_{\alpha}-\orr_{C})/a]}
\eea
with $\orr_{C}=1.2A_D^{1/3}$ the standard geometrical contact
radius and the nuclear diffusivity $a=0.7$.

\section{Applications}
\label{numap}

\begin{figure}
\centering 
\includegraphics[width=0.4\textwidth]{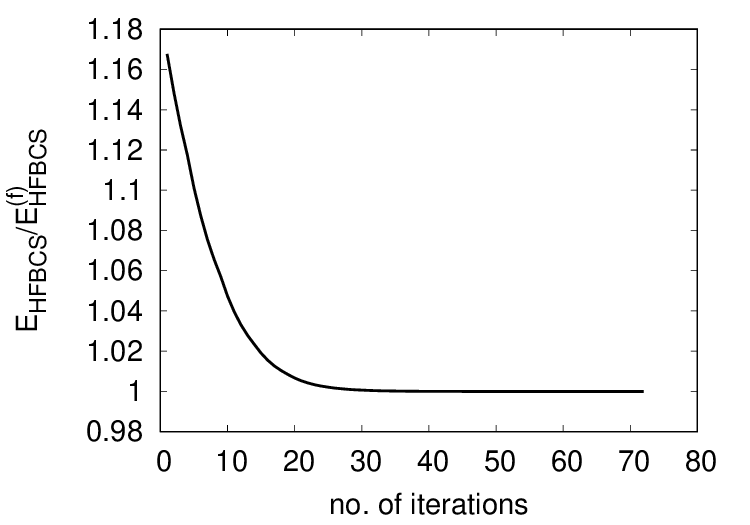}	
\caption{Ratio between the current and final values of the gs energy
$E_{\textrm{HFBCS}}$ as function of the number of iterations for 
$\tensor[^{242}]{\textrm{Pu}}{}$. In this example, the decay width is 
reproduced by involving $3$ clusters in the antisymmetrization.} 
\label{fig1}
\end{figure}

The theory previously developed is put to practice as follows. Starting
from an axially-symmetric MF configuration given by a standard WS field
in the universal parameterization \cite{Dud82,Cwi87}, one first 
constructs a SCMF by calculating radial form factors of a potential
according to Eq. (\ref{uuffs}) and merging this result with the 
previous one via the rule 

\bea
\label{merge}
U_{L}^{\lp\textrm{new}\rp}\lp r\rp=\lp 1-\xi\rp 
U_{L}^{\lp\textrm{old}\rp}\lp r\rp+
\xi U_{L}^{\lp\textrm{calc}\rp}\lp r\rp
\eea
where $\xi$ starts at the value $0.1$ and gradually tends to $1$ as
convergence is approached. For nuclei above Pb this is typically
achieved in 70--80 iterations, each step involving a MF diagonalization
followed by a BCS calculation. Once convergence has been reached, one
calculates the $\alpha$--particle formation amplitude. The calculation
is redone for different sets of the clustering strengths $x_{\tau}$ 
until a good description of both the nucleus and decay width have been
achieved. The only time--consuming part of the procedure is the
initial evaluation of the $v_{L}^{\lp\textrm{com}\rp}$ 
quantities. The choice of effective lengths for the rel and com parts is 
universally fixed at $b=B=1~\textrm{fm}$. This choice for $b$ strikes a 
compromise between realism and an expeditious evaluation of 
$v_{L}^{\lp\textrm{com}\rp}$. For $B$, the same value
has been used throughout our previous studies and it was found to 
have a small influence on the bulk of the sp spectrum while also giving
good results for the $\alpha$--particle formation amplitude. It should
be mentioned in the context of today's computational developments that 
the calculation of $v_{L}^{\lp\textrm{com}\rp}$ can be an interesting
application of machine learning (ML) software such as TensorFlow 
\cite{Aba16}. However, for this work we ultimately resorted to classical
integration methods that were sped up via parallel computing techniques 
\cite{Dag98}. The isospin--dependent interaction strength $\ov_{\tau}$
is fixed by setting $x_{\tau}=0$ and then minimizing the deviation 
between the starting MF and the calculated direct term. The value varies
slightly with the prescribed elongation $\beta_{2}$, so for each nucleus
we determined these constants over a range $\beta_{2}\in\lb-0.4,0.4\rb$
and then used the average value. In practical calculations, one can use
a value for $\beta_{2}$ as prescribed in liquid drop models (LDM) 
\cite{Mol95} or the value corresponding to the HFBCS energy minimum, as 
we will show below. The spherical factor of the clustering radius has
been taken from the systematics \cite{Dum22} in terms of the daughter
nuclear mass $A_{D}$
\bea
\orr_{0}=1.168\lp A_{D}^{\frac{1}{3}}+1.584\rp
\eea
which is fairly close to the typical geometrical contact radius 
$\orr_{C}$ and to the Mott point where nucleonic density drops to about 
$10\%$ of its central value \cite{Rop98}. It should be noted however 
that the final clustering radius within the SCMF will be found slightly 
more inward, due to the integration involved in the evaluation of 
$U_{L}$. Furthermore, as was discussed in detail in our previous 
investigations \cite{Dum25}, two--particle interactions of the form 
(\ref{vsgi}) tend to be unstable in HF calculations, leading either to a 
collapse of the nucleus or an unphysical distribution of nucleons. We 
have corrected this in a manner similar to the one used previously, 
namely by multiplying the integrand in Eq. (\ref{uuffs}) for the 
monopole term with the ratio 
$\frac{\rho_{0}\lp r\rp}{\langle\rho^{\lp i\rp}\lp r\rp\rangle}$.
At any given point, the numerator is just the value of the original MF
density while the denominator is an average taken over a small 
neighborhood around that point for the density in that given iteration.
One cannot use the same approach for the $L=2$ quadrupole term due to 
the fact that such terms typically have nodes near the origin. However,
the quadrupole terms can be fitted quite well overall with gaussian 
functions and we have used such gaussian fits in the merging process
(\ref{merge}) for $L=2$. We have observed that these two corrections 
cure all pathological behaviors and stabilize the HFBCS process. As
an illustration, we consider the case of the emitter 
$\tensor[^{242}]{\textrm{Pu}}{}$. We define the HFBCS energy as
\bea
E_{\textrm{HFBCS}}=2\sumd_{m>0}\epsilon_{m}v_{m}^{2}-
\frac{\Delta^{2}}{G}
\eea
for both isospins in terms of the BCS  energy gap $\Delta$ and the 
constant strength $G$. Fig. \ref{fig1} shows the ratio of the current
$E_{\textrm{HFBCS}}$ value to the final one versus the number of 
iterations for the case where the decay width is reproduced considering
that $3$ clusters participate in the antisymmetrization. We observe
that the process is stabilized in roughly $30$ iterations, with only
small adjustments following that point.

\begin{figure}
\centering 
\includegraphics[width=0.4\textwidth]{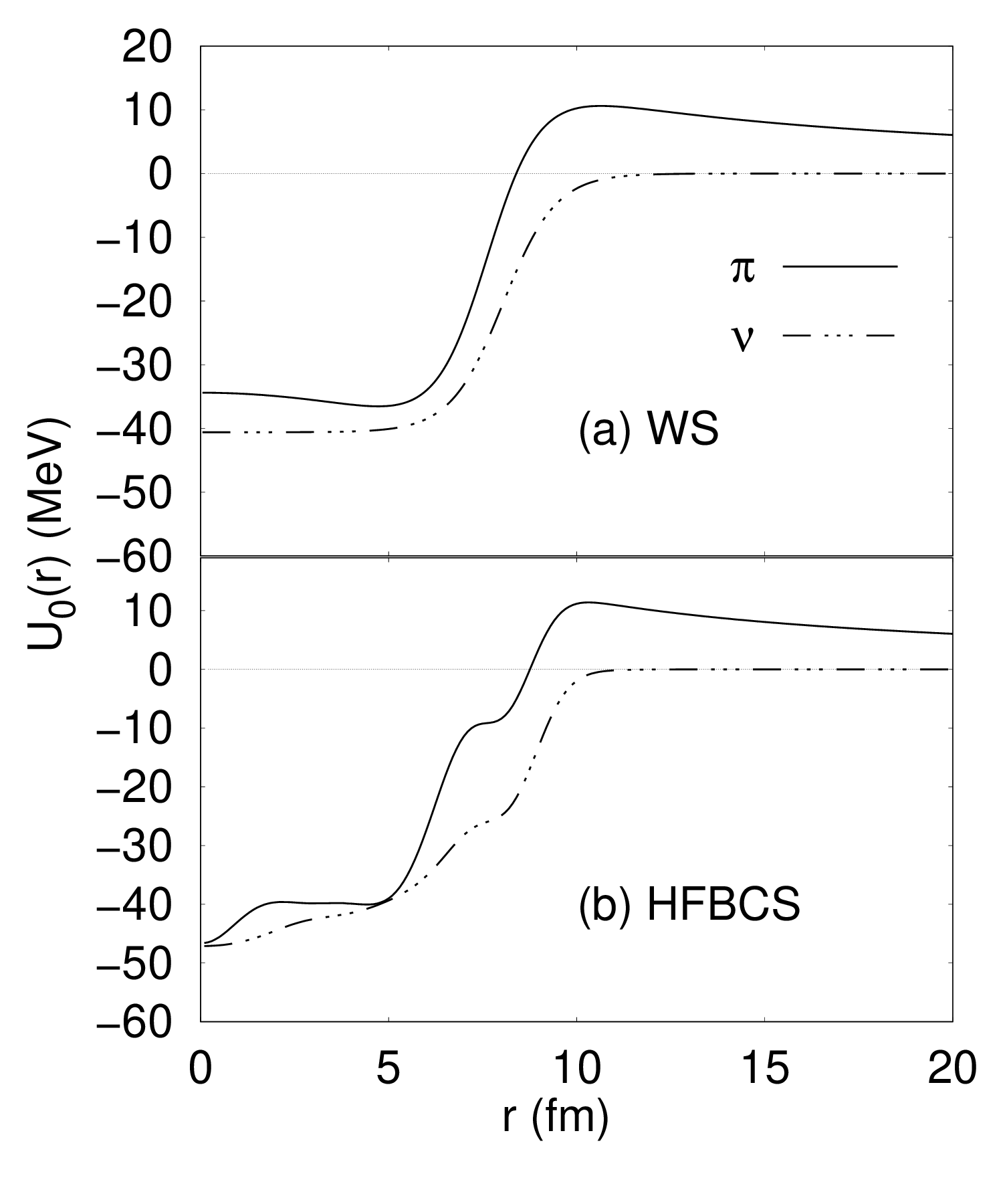}	
\caption{Initial proton (continuous line) and neutron (dot--dashed line) 
WS field (a), and HFBCS field (b) when the decay width
is reproduced for two clusters participating in the antisymmetrization.} 
\label{fig2}
\end{figure}

As an example of a highly clustered configuration, we look at the decay
of $\tensor[^{216}]{\textrm{Rn}}{}$. In Fig. \ref{fig2} panel (a)
we show the inital WS plus Coulomb monopole fields for protons
and neutrons versus radius. Panel (b) shows the result of our HFBCS 
calculation when the decay width is reproduced, considering that two 
clusters participate in the antisymmetrization. In both fields, we 
observe the formation of surface pockets that locally enhance the 
nucleonic density and increase the contribution of the clustered 
configuration to the gs.

\begin{figure}
\centering 
\includegraphics[width=0.4\textwidth]{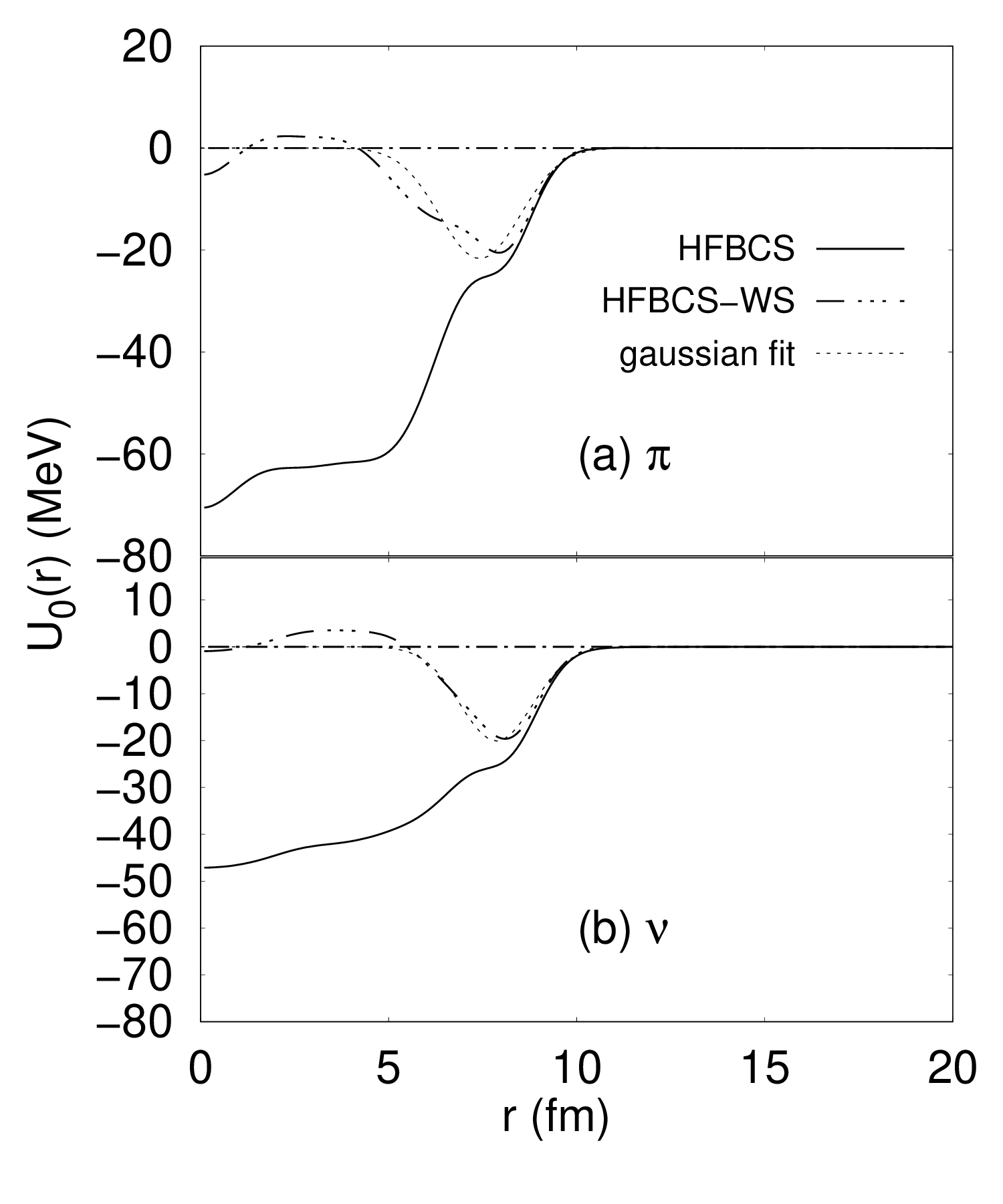}	
\caption{HFBCS fields, the surface components after removal of the WS
part and the fitting surface gaussian for the monopole term in the case
of protons (a) and neutrons (b).} 
\label{fig3}
\end{figure}

\begin{figure}
\centering 
\includegraphics[width=0.4\textwidth]{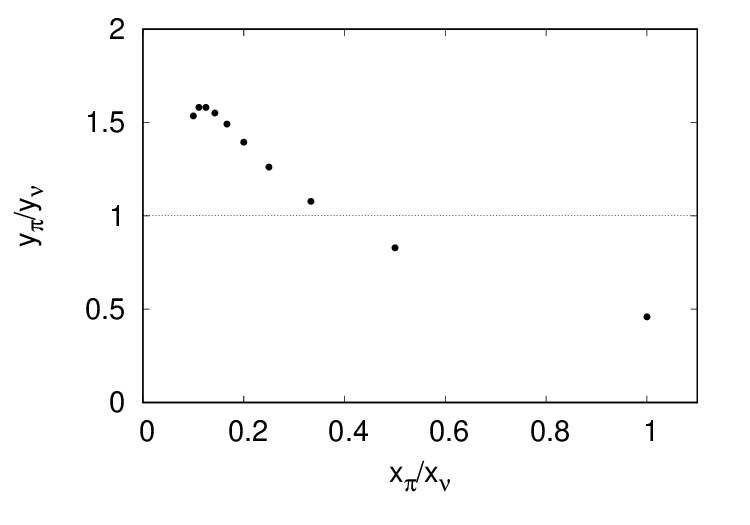}	
\caption{Ratio of proton to neutron surface amplitudes in the SCMF 
versus the same ratio of the surface amplitudes in the two--particle
interaction for $\tensor[^{216}]{\textrm{Rn}}{}$.}
\label{fig4}
\end{figure}

To understand the effect of $v_{L}^{\lp\textrm{com}\rp}$ on the 
resulting SCMF, we plot once more the results in the case of 
$\tensor[^{216}]{\textrm{Rn}}{}$ in Fig. \ref{fig3}, separately for
protons and neutrons, without the Coulomb term. The continuous lines 
represent the HFBCS fields for the clustered configuration. The 
dot--dashed line is the HFBCS field from which we have extracted the 
best fit that would be provided by a WS potential. The result is very 
well approximated by a surface gaussian, depicted by a dotted line. 
Given that the $\alpha$--clusters are formed in these pockets found in 
the internal region, one would expect the physical result to be isospin 
invariant, namely the proton and neutron pockets should have similar 
shapes. However, the amplitude $x_{\tau}$ within the interaction
cannot be the same for both particles, due to the presence of the 
Coulomb barrier. In Fig. \ref{fig4} we plot the ratio of the resulting 
amplitudes $y_{\tau}$ for the surface corrections in the SCMF as a 
function of the ratio of amplitudes $x_{\tau}$ entering the 
two--particle interaction. The calculations are done for 
$\tensor[^{216}]{\textrm{Rn}}{}$, but the results are valid for all 
emitters we have investigated above $\tensor[^{208}]{\textrm{Pb}}{}$. 
Namely, by taking the ratio of the initial proton to neutron amplitudes 
$x_{\tau}$ as $\frac{1}{3}$, one obtains surface pockets of nearly 
identical amplitudes in the corresponding SCMFs. We note that in the 
case of Xe the smaller Coulomb barriers require the ratio to be about 
$0.91$.

\begin{figure}
\centering 
\includegraphics[width=0.4\textwidth]{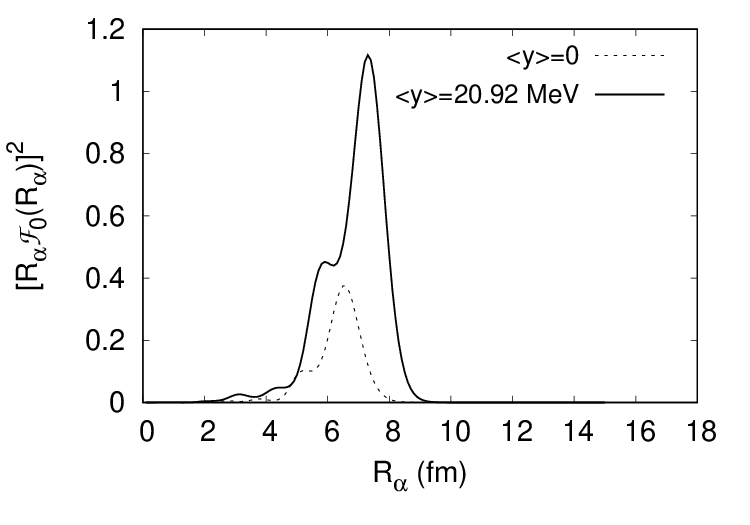}	
\caption{$\alpha$--particle formation amplitude versus the 
$\alpha$--particle com for $\tensor[^{216}]{\textrm{Rn}}{}$ when no 
clustering is present (dashed line) and with surface corrections 
(continous line) in the SCMF having average amplitudes of about 21 MeV, 
which reproduce the experimental decay width.}
\label{fig5}
\end{figure}

\begin{figure}
\centering 
\includegraphics[width=0.4\textwidth]{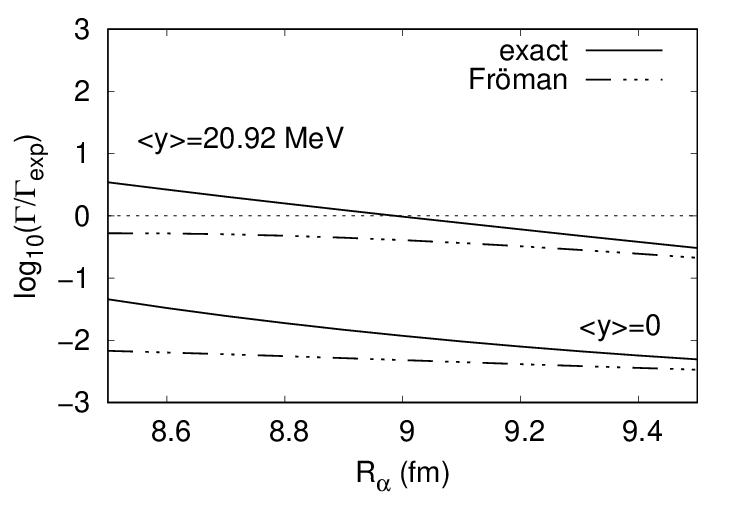}	
\caption{Logarithm of the ratio between theoretical and experimental
decay widths for $\tensor[^{216}]{\textrm{Rn}}{}$ calculated
exactly (continuous line) and in the Fr\"{o}man approximation
(dot--dashed line) with and without surface clustering and including
two clusters in the antisymmetrization.}
\label{fig6}
\end{figure}

The effect of the surface term on the $\alpha$--particle formation 
amplitude as function of the $\alpha$--particle's com for the decay of
$\tensor[^{216}]{\textrm{Rn}}{}$ is shown in Fig. \ref{fig5}. The dashed 
line represents a HFBCS result with the surface corrections set to 0 and 
two clusters present in the antisymmetrization, a situation where the 
experimental decay width is underestimated by almost two orders of 
magnitude. The continuous line represents the same HFBCS calculation, 
but with an average surface amplitude of the correction terms of roughly 
21 MeV. The formation amplitude itself is then greatly enhanced and its 
peak is slightly shifted towards smaller densities. With these two cases 
in mind, we plot the logarithm of the ratio of the calculated to 
experimental decay width as a function of the $\alpha$--particle com in 
Fig. \ref{fig6}. The continuous lines represent exact calculations with 
and without clustering, while the dot--dashed lines are the same 
calculations carried out in the Fr\"{o}man approximation. We observe 
that the Fr\"{o}man calculation slightly underestimates the experimental 
result when averaged around 9 fm, the geometrical contact radius for the 
decay of $\tensor[^{216}]{\textrm{Rn}}{}$, while the exact calculation 
reproduces the observed decay width when clustering is included in the
two--particle interaction. All calculations are done with two clusters
present in the antisymmetrization.

\begin{figure}
\centering 
\includegraphics[width=0.4\textwidth]{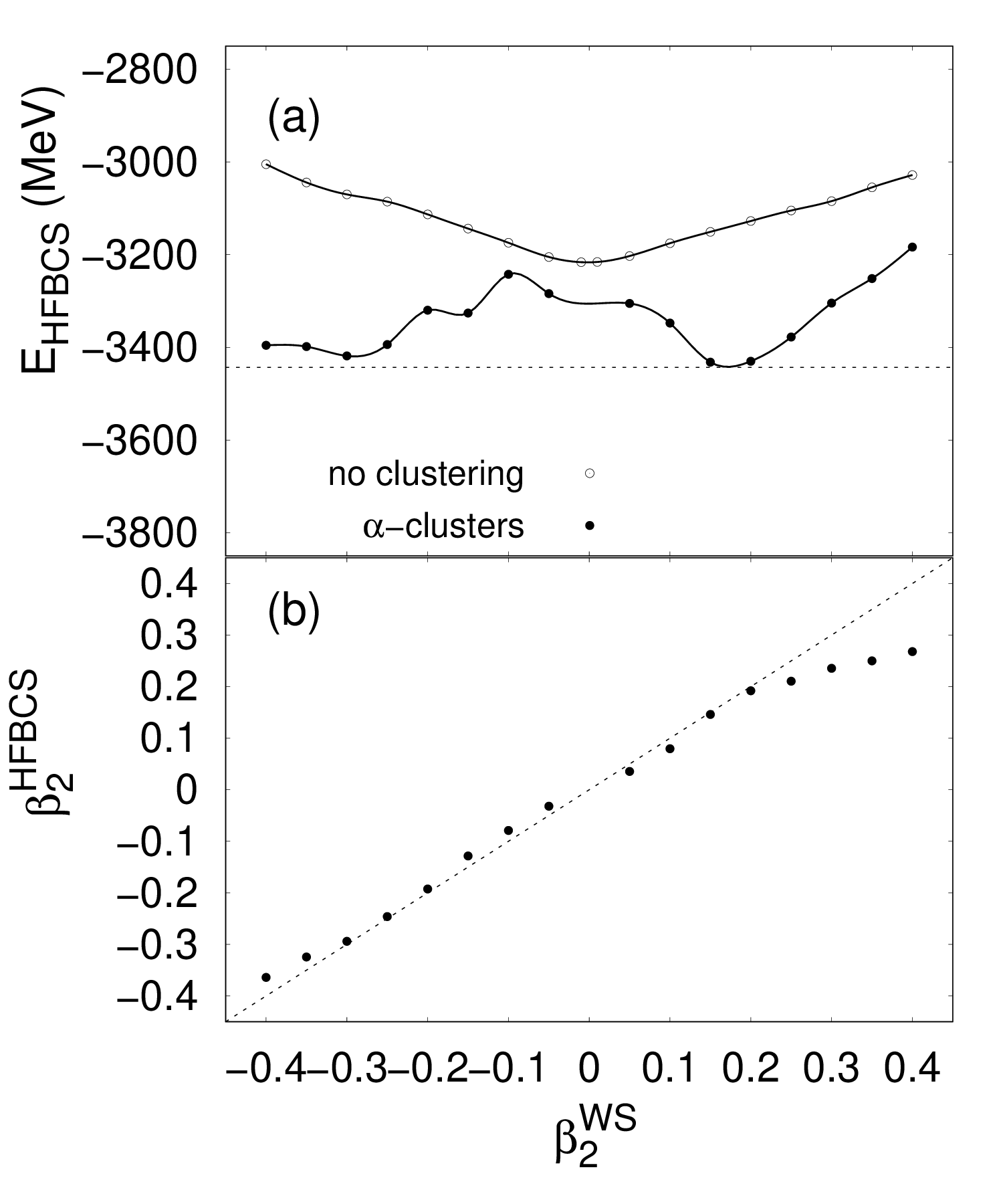}	
\caption{gs HFBCS energy as a function of the WS quadrupole
elongation (a) for no clustering (open circles) and with enough 
clustering in the interaction in order to reproduce the experimental 
decay width (dark circles) of $\tensor[^{242}]{\textrm{Pu}}{}$. 
Panel (b) shows the corresponding quadrupole elongations of the HFBCS 
clustered configurations as a function of the initial WS elongation.}
\label{fig7}
\end{figure}

The formation of an $\alpha$--particle involves an interplay between
antisymmetrization, clustering and the nuclear deviation from a 
spherical shape. To see this, we first look at how the gs energy and 
nuclear shape vary as a function of the quadrupole elongation, with and
without $\alpha$--clusters present in the final HFBCS configuration. 
This is shown in Fig. \ref{fig7} for the decay of 
$\tensor[^{242}]{\textrm{Pu}}{}$. Panel (a) represents the gs HFBCS 
energy versus the initial elongation of the WS field in two cases. The 
open circles are calculations done with no clustering component present
in the two--particle interaction, where we observe that the equilibrium
shape is a spherical one. The dark circles represent the same 
calculation but carried out with the appropriate amount of clustering
that reproduces the observed decay width. Overall, the energies are
lower and two deformed minima appear corresponding to oblate and prolate
shapes, the prolate minimum being slightly deeper. It is also fairly 
close to the LDM value $\beta_2=0.215$ \cite{Mol95} for the daughter 
nucleus $\tensor[^{238}]{\textrm{U}}{}$. This is an interesting result, 
as the deformed minima were obtained from a different kind of 
constraint, namely fixing the decay width of the gs to its experimental 
value, as opposed to the usual quadrupole moment cranking performed in 
HFB calculations. Panel (b) shows the final nuclear elongation for the 
HFBCS state versus the initial elongation, each point corresponding to 
the calculation where the decay width is reproduced. We see that the 
final deformations remain fairly close to their original values for the 
oblate case, while prolate shapes tend to become less deformed as the 
input elongation increases.

\begin{figure}
\centering 
\includegraphics[width=0.4\textwidth]{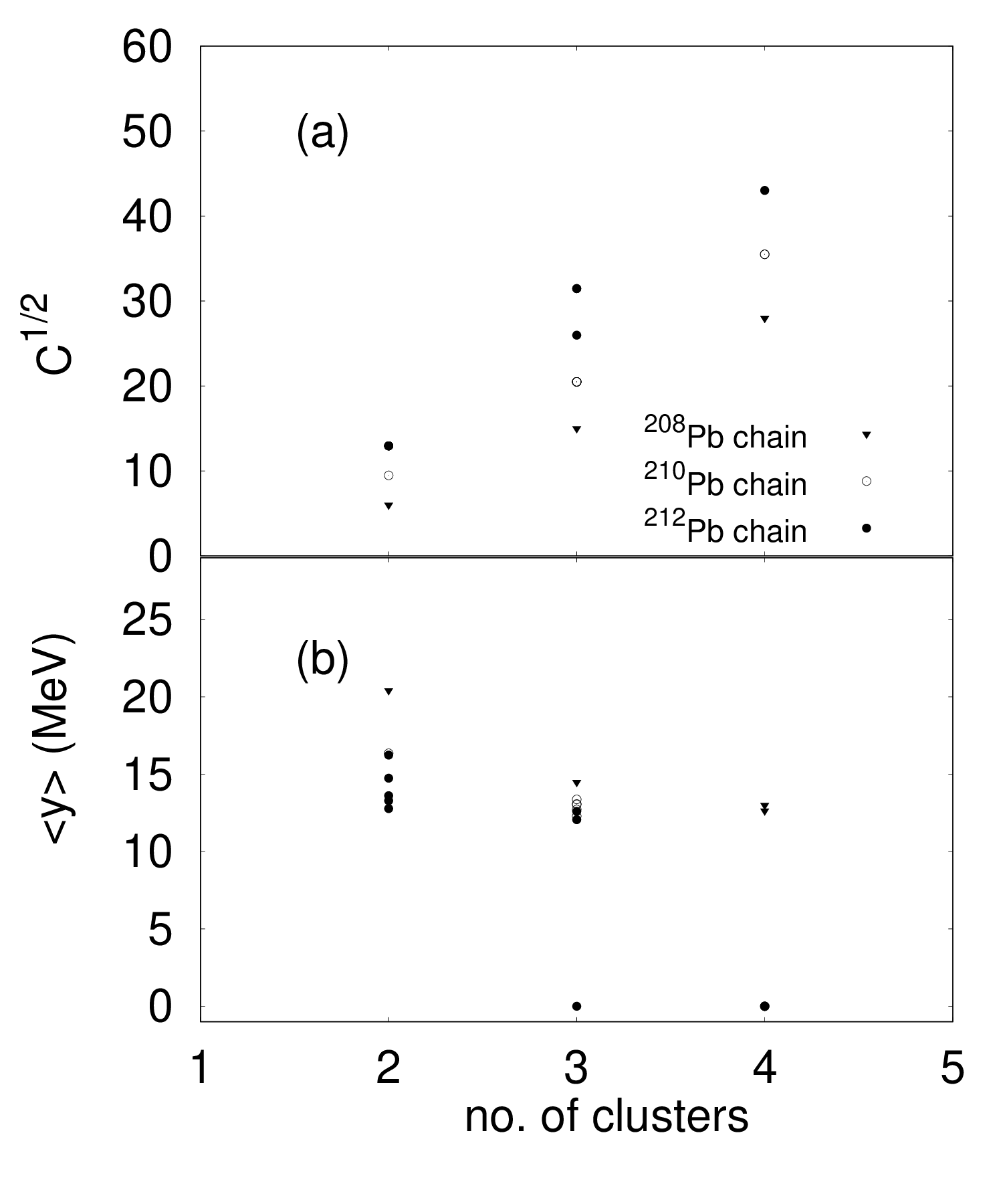}	
\caption{Binomial coeficient entering the antisymmetrization formula
as a function of the number of clusters (a) for decay chains terminating
in $\tensor[^{208,210,212}]{\textrm{Pb}}{}$. The corresponding average
surface cluster required to reproduce the experimental decay width
as a function of the number of clusters (b). }
\label{fig8}
\end{figure}

\begin{figure}
\centering 
\includegraphics[width=0.4\textwidth]{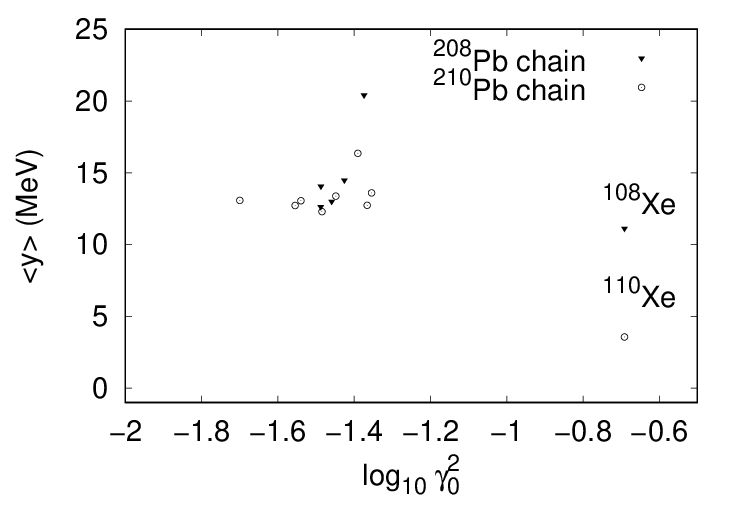}	
\caption{Average surface clustering in the SCMF for the decay chains of
$\tensor[^{208,210}]{\textrm{Pb}}{}$ as a function of the reduced decay
width. The data for $\tensor[^{108,110}]{\textrm{Xe}}{}$ are also 
represented with the same corresponding symbols as the decay chains.}
\label{fig9}
\end{figure}

In Fig. \ref{fig8} we show the results for calculations performed along
decay chains terminating in $\tensor[^{208,210,212}]{\textrm{Pb}}{}$,
corresponding to dark triangles, open circles and filled circles. We 
start the calculations from the appropriate Rn parents and increment the 
mass number by an $\alpha$--particle. Panel (a) shows the square root of 
the binomial coefficient $C$ entering the antisymmetrization formula 
as a function of the number of clusters allowed before the experimental
decay width is obtained. For the heavier chains, all excess neutrons are 
taken along with the clusters. Panel (b) shows the corresponding average
surface cluster amplitudes required in the SCMF so that the experimental
decay width is well reproduced, the vanishing values corresponding to
experimental widths reproduced just from antisymmetrization. The inverse
correlation is very clear, a larger antisymmetrization requiring a 
smaller cluster component in the interaction (and consequently, a 
smaller cluster amplitude on the nuclear surface) in order for the 
experimental width to be reproduced. Fig. \ref{fig9} shows the same
averaged clustering amplitudes for the 
$\tensor[^{208,210}]{\textrm{Pb}}{}$ decay chains versus the 
experimental reduced width calculated from the standard formula
\bea
\log_{10}\Gamma=\log_{10} 2P\lp\orr_{0}\rp+
\log_{10}\gamma_{0}\lp\orr_{0}\rp
\eea
where $P$ is the usual penetrability function. Along with these values,
for the $\tensor[^{208}]{\textrm{Pb}}{}$ chain we added 
$\tensor[^{108}]{\textrm{Xe}}{}$ (dark triangles) and for
$\tensor[^{210}]{\textrm{Pb}}{}$ we added 
$\tensor[^{110}]{\textrm{Xe}}{}$ (open circles). We see that though 
the experimental reduced width is larger for Xe isotopes, the surface
clustering required to obtain it is smaller than for the corresponding
Pb decay chain, due to the presence of antisymmetrization. We note that
for calculations performed without antisymmetrization in the same region
above $\tensor[^{100}]{\textrm{Sn}}{}$, a stronger cluster component 
than the one found above $\tensor[^{208}]{\textrm{Pb}}{}$ is required
in order to reproduce the experimental decay widths \cite{Bar16}.

\section{Summary and conclusions}
\label{conc}

We have presented a self--consistent field calculation for 
axially--symmetric nuclei constructed from a two--particle interaction
having rel and com terms, the com term serving as an enhancement of 
nucleonic clustering on the surface of the nucleus. We have developed
a corresponding cluster formation theory for the axially-symmetric 
system thus described and we have proposed an approximate treatment of
the daughter nucleus--cluster antisymmetrization. Applications have
concerned mainly deformed $\alpha$--emitters above  
$\tensor[^{208}]{\textrm{Pb}}{}$ together with some examples above
$\tensor[^{100}]{\textrm{Sn}}{}$. We have seen that there is an 
interplay between antisymmetrization and the surface clustering of 
nucleons, typically both being required in order for the 
$\alpha$--particle to form near the Mott point and decay according 
to its experimental width. We have also seen how these aspects of the
decay process affect the final shape of the nucleus and have used the
experimental decay width as a constraint forcing a deformed minimum
in the gs energy as a function of the nuclear elongation. We have
noted that the presence of antisymmetrization diminishes the required
clustering component in order for the decay width to be reproduced
in emitters above $\tensor[^{100}]{\textrm{Sn}}{}$, a result at
variance with the theoretical conclusions that predict enhanced
clustering in this region.

\section*{Acknowledgements}

This work was supported by a grant of the Ministry of Research, 
Innovation and Digitization, CNCS--UEFISCDI, Project No. 
PN--IV--P1--PCE--2023--0273, within PNCDI IV, and from the grant having
Project No. PN--23--21--01--01/2023.

\vspace{2pc}

\bibliographystyle{vancouver} 
\bibliography{nuclear}

\begin{thebibliography}{10}

\bibitem{Gam28}
Gamow G.
\newblock Zur Quantentheorie des Atomkernes.
\newblock Zeitschrift f\"{u}r Physik. 1928;51(3-4):204-12.

\bibitem{Gur28}
Gurney RW, Condon EU.
\newblock Wave Mechanics and Radioactive Disintegration.
\newblock Nature. 1928;122:439.

\bibitem{Lov98}
Lovas RG, Liotta RJ, Insolia A, Delion DS.
\newblock Microscopic theory of cluster radioactivity.
\newblock Physics Reports. 1998;294(5):265-362.

\bibitem{Del15}
Delion DS, Dumitrescu A.
\newblock Systematics of the $\alpha$--decay fine structure in even--even
  nuclei.
\newblock Atomic Data and Nuclear Data Tables. 2015;101:1.

\bibitem{Del18}
Delion DS, Ren Z, Dumitrescu A, Ni D.
\newblock Coupled channels description of the $\alpha$--decay fine structure.
\newblock Journal of Physics G: Nuclear and Particle Physics. 2018;45(5).

\bibitem{Dum22}
Dumitrescu A, Delion DS.
\newblock The phenomenology of particle and cluster emission.
\newblock Atomic Data and Nuclear Data Tables. 2022;145:101501.

\bibitem{Dum25}
Dumitrescu A, Delion DS.
\newblock Emission processes in a self--consistent field.
\newblock Journal of Physics G: Nuclear and Particle Physics.
  2025;52(6):055107.

\bibitem{San62}
Sandulescu A.
\newblock Reduced widths for favoured alpha transitions.
\newblock Nuclear Physics. 1962;37:332-43.

\bibitem{Man64}
Mang HJ.
\newblock Alpha Decay.
\newblock Annual Review of Nuclear and Particle Science. 1964;14:1-26.

\bibitem{Pog69}
Poggenburg JK, Mang HJ, O RJ.
\newblock Theoretical Alpha--Decay Rates for the Actinide Region.
\newblock Physical Review. 1969;181(4).

\bibitem{Lan60}
Lane AM.
\newblock Reduced Widths of Individual Nuclear Energy Levels.
\newblock Reviews of Modern Physics. 1960;32(3).

\bibitem{Fli75}
Fliessbach T.
\newblock The reduced width amplitude in the reaction theory for composite
  particles.
\newblock Zeitschrift f\"{u}r Physik A Atoms and Nuclei. 1975;272:39-46.

\bibitem{Fli76}
Fliessbach T, Mang HJ, Rasmussen JO.
\newblock Normalized shell model alpha decay theory applied to unfavored decay.
\newblock Physical Review C. 1976;13(3).

\bibitem{Fli77}
Fliessbach T.
\newblock A Model Calculation of the Aymptotic Reduced $\alpha$--Amplitude.
\newblock Nuclear Physics A. 1977;285:262-8.

\bibitem{Fli79}
Fliessbach T, Manakos P.
\newblock The Reduced $\alpha$--Amplitude in an Exactly Solvable Model.
\newblock Nuclear Physics A. 1979;324:173-81.

\bibitem{Rop98}
R\"opke G, Schnell A, Schuck P, Nozi\`eres P.
\newblock Four-Particle Condensate in Strongly Coupled Fermion Systems.
\newblock Physical Review Letters. 1998;80(15):3177-80.

\bibitem{Toh01}
Tohsaki A, Horiuchi H, Schuck P, R\"{o}pke G.
\newblock Alpha Cluster Condensation in $^{12}\textrm{C}$ and
  $^{16}\textrm{O}$.
\newblock Physical Review Letters. 2001;87:192501.

\bibitem{Epe11}
Epelbaum E, H K, Lee D, Mei\ss{}ner UG.
\newblock \textit{Ab Initio} Calculation of the Hoyle State.
\newblock Physical Review Letters. 2011;106:192501.

\bibitem{Ebr12}
Ebran JP, Khan E, Nik\ifmmode \check{s}\else \v{s}\fi{}i\ifmmode~\acute{c}\else
  \'{c}\fi{} T, Vretenar D.
\newblock How atomic nuclei cluster.
\newblock Nature. 12012;487:341-3.

\bibitem{Das23}
Dassie AC, Id~Betan RM.
\newblock $\alpha$--decay from {$^{44}$Ti}: A study of microscopic
  clusterization.
\newblock Physical Review C. 2023;108:044314.

\bibitem{Bet12}
Id~Betan R, Nazarewicz W.
\newblock $\alpha$ decay in the complex energy shell model.
\newblock Physical Review C. 2012;86(3):034338.

\bibitem{Rop14}
R\"opke G, Schuck P, Funaki Y, Horiuchi H, Ren Z, Tohsaki A, et~al.
\newblock Nuclear clusters bound to doubly magic nuclei: The case of
  $^{212}\mathrm{Po}$.
\newblock Phys Rev C. 2014;90(3):034304.

\bibitem{Car21}
Carstea AS, Ludu A.
\newblock Nonlinear Schr\"{o}dinger equation solitons on quantum droplets.
\newblock Physical Review Research. 2021;3(3):033054.

\bibitem{Del13}
Delion DS, Liotta RJ.
\newblock Shell-model representation to describe $\alpha$ emission.
\newblock Physical Review C. 2013;87(4):041302.

\bibitem{Dum23}
Dumitrescu A, Delion DS.
\newblock Cluster mean-field description of $\alpha$ emission.
\newblock Physical Review C. 2023;107(2):024302.

\bibitem{Del10}
Delion DS.
\newblock Theory of Particle and Cluster Emission.
\newblock Springer Nature; 2010.

\bibitem{Dob21}
Dobaczewski J, Baczyk P, Becker P, et~al.
\newblock Solution of universal nonrelativistic nuclear {DFT} equations in the
  {Cartesian} deformed harmonic-oscillator basis. ({IX}) {HFODD} (v3.06h): a
  new version of the program.
\newblock Journal of Physics G: Nuclear and Particle Physics.
  2021;48(10):102001.

\bibitem{Fro57}
Fr{\"{o}}man PO.
\newblock Alpha Decay of Deformed Nuclei.
\newblock Mat Fys Skr Dan Vid Selsk. 1957;1(3):1-74.

\bibitem{Dud82}
Dudek J, Szymanski Z, Werner T, Faessler A, Lima C.
\newblock Description of high spin states in $^{146}$Gd using the optimized
  Woods--Saxon potential.
\newblock Physical Review C. 1982;26:1712-8.

\bibitem{Cwi87}
Cwiok S, Dudek J, Nazarewicz W, Skalski J, Werner T.
\newblock Single--particle energies, wave functions, quadrupole moments and
  g--factors in an axially deformed Woods--Saxon potential with applications to
  the two--center--type nuclear problem.
\newblock Computer Physics Communications. 1987;46:379-99.

\bibitem{Aba16}
Abadi M, P B, et~al CJ.
\newblock TensorFlow: A System for Large--Scale Machine Learning.
\newblock Proceedings of the 12th USENIX Symposium on Operating Systems Design
  and Implementation (OSDI '16). 2016.

\bibitem{Dag98}
Dagum L, Menon R.
\newblock OpenMP: an industry standard API for shared-memory programming.
\newblock IEEE Computational Science and Engineering. 1998;5(1):46-55.

\bibitem{Mol95}
Moller P, Nix JR, Myers WD, Swiatecki WJ.
\newblock Nuclear Ground--State Masses and Deformations.
\newblock Atomic Data and Nuclear Data Tables. 1995;59(2):185-381.

\bibitem{Bar16}
Baran VV, Delion DS.
\newblock Proton-neutron versus $\ensuremath{\alpha}$-like correlations above
  $^{100}\mathbf{Sn}$.
\newblock Phys Rev C. 2016 Sep;94:034319.

\end{thebibliography}

\end{document}